# Optoelectronic and non-linear optical properties of Lu-doped AgGaGe3Se8 Crystallites

V. Kityk1,4 · G. L. Myronchuk1 · M. Lelonek2 · P. Goring2 · L. Piskach3 · B. Vidrynsky1 · A. Ryzhuk1 · A. O. Fedorchuk5 · J. Jedryka4

G. L. Myronchuk
g_muronchuk@ukr.net
I. V. Kityk
iwank74@gmail.com

1 Department of Experimental Physics and Information‑Measuring Technology, Lesya Ukrainka Eastern European National University, Voli Avenue 13, Lutsk 43025, Ukraine
2 SmartMembranes GmbH, Heinrich‑Damerow‑Str. 4, 06120 Halle, Germany
3 Department of Inorganic and Physical Chemistry, Lesya Ukrainka Eastern European National University, 13 Voli Avenue, 43025 Lutsk, Ukraine
4 Institute of Optoelectronics and Measuring Systems, Faculty of Electrical Engineering, Czestochowa University of Technology, Armii Krajowej 17 Str., 42‑200 Czestochowa, Poland
5 Department of Inorganic and Organic Chemistry, Lviv National University of Veterinary Medicine and Biotechnologies, 50 Pekarska St., Lviv 79010, Ukraine

**Abstract**
A complex studies of optoelectronics, non-linear optical and laser stimulated piezo electric features of chalcogenide powder-like chalcogenide crystals pure and rare earth moped AgGaGe3Se8 crystallites are presented. It is shown principal role of the morphology for the titled materials. The contribution to nonlinear optical, photoconductivity and laser stimulated piezoelectricity are comparable. The temperature dependences are explored. The possibilityity to operate by the features of the titled crystallites in the optically reflected regime is shown. This may be important for laser operated triggers, modulators, photo detectors etc. The relaxation processes are studied. Among the NLO features main attention is devoted to second harmonic generation efficiencies in the reflected regime where principal role is played by morphology. Additionally the laser stimulated piezoelectric for the near the surface states with different morphology of grains is explored. It is demonstrated that contribution of the near-the surface states for such kind of effects will be commensurable to the bulk-like contribution.

## 1 Introduction

Recently, one can observe an enhanced interest in the design of IR optically triggered and operated devices based on chalcogenides. Particular interest presents the ternary and quaternary crystals (Kityk et al. 2004; Bi et al. 2007; Kulyk et al. 2009; Sahraoui et al. 2010; Iliopoulos et al. 2013). Chalcogenide crystals, unlike oxides, have larger transparenty region that covers also the mid-IR spectral range (Liang et al. 2017). During transition from S to Se to Te halogens become more polarizable, the band gap energy decreases, and nonlinear optical susceptibilities both of the second as well as third order are enhanced. At the same time incorporation of rare earth atoms as a rule leads to significant decrease of effective energy gap and decrease of transparency which is the principal parameter for optoelectronic applications. For the future development of effective multifunctional IR laser operated devices (frequency multiplied modulators, deflectors etc). In the present work we will focus particular attention on the study of promising chalcogenide non-linear optical crystals doped with rare earth elements. Moreover, most of the studies are performed for the bulk crystals, however due to a huge number of instrisc defects and imperfections very important to explore their use in the powder like form. Here very crucial role will be player by the grain morphology which is usually not considered during the investigation. The main problem with multi-component chalcogenide materials used in optoelectronic devices is the spatial optical non-homogeneity due to the large number of intrinsic cation defects. The consideration of the near-the surface morphology is closely related to the occurrence the low dimensional (micrometer and nano-meter sheets) which will essentially change the optoelectronic features On the one hand, the such reconstructed morphology also promotes photoinduced properties, and on the other, they lead to additional photo-thermal effects that also change the principal dipole moments during laser illumination. Another import ant factor here is a large phonon anharmonisms (Shpotyuk et al. 1997) in chalcogenides that lead to the appearance of large non-linear optical effects. The most popular crystalline chalcogenide materials for the IR region are such materials as $AgGaS_2$, $AgGaSe_2$ (Abrahams and Bernstein 1973). The solid solution range of $AgGaGe_3Se_8$ is simialr to $AgGaSe_2$. They were found in the $AgGaSe_2$–$GeSe_2$ system in the work to improve non-linear optical and photoelectric parameters of the ternary phases by adding germanium dichalcogenides (Olekseyuk et al. 2002). Addition of $GeS_2$ to $AgGaS_2$ and $GeSe_2$ to $AgGaSe_2$ was found to enhance three principal parameters: birefringence, band gap magnitude, and radiation resistance (Badikov et al. 2009). Birefringence is double from 0.05 for $AgGaSe_2$ to 0.11 for $AgGaGe_3Se_8$ which greatly extends the spectra range of operating wavelength. Improved laser damage threshold of the quaternary crystals makes them also a promising alternative to the widelyup used $AgGaS_2$ and $AgGaSe_2$ for the Nd:YAG laser frequency converter, as well as for many other applications (Petrov et al. 2004).

## 2 Experimental

The crystals were synthesized by the traditional Bridgman-Stockbarger method. The conditions for the single crystal growth were as follows: temperature in the crystallization band—1250 K; annealing temperature—720 K; the temperature gradient at the solid-melt interface—5 K/mm; growth rate—0.1 mm/h; annealing time—150 h; rate of cooling to ambient—5 K/h. The $AgGaGe_3Se_8$ single crystals thus obtained had forms of cylinder with 18 mm in diameter and 30 mm in length. Lu impurity (about 0.2%) was introduced into the starting batch to obtain concentration close to the concentration of intrinsic structural defects. X-ray diffraction (XRD) structural studies have shown that Lu doping did not alter the structure or the basic lattice parameters of the $AgGaGe_3Se_8$ crystal. However, there was a noticeable change in the height and width of the diffraction reflections. The observed transformation of the XRD patterns indicates the distortions caused by the admixture and the appearance of the stressed state of the matrix lattice. Infrared transmission spectra were investigated using a FTIR Spectrum Two ™ spectrophotometer (PerkinElmer). The optical system with KBr windows allows to perform the measurements in the spectral range 7800–370 $cm^{-1}$ with 0.5 $cm^{-1}$ resolution. Spectral and space distribution of the absorption coefficient in the region of the fundamental absorption edge band was also explored. Parallelepiped plates of 0.06–0.1 mm thickness were prepared of Ag-containing crystals for the measurements. The plates were polished in castor oil containing ultra-fine abrasive additives and crushed using external mechanical and acoustical fields (average particle size ~ 28 μm). These powders have been used for different researches. An MDR-206 diffraction monochromator with silicon photodiode was used as a spectral instrument in the spectral range 360–1100 nm (spectral resolution 0.2 nm). Temperature-dependent studies were performed in thermo regulated helium and nitrogen cryostat with thermo stabilization up to

0.02 K. Another important parameter which is determined by the morphology is piezo electric effect and possibility of its changes under influence of external laser light Piezoelectric studies were performed using piezoelectric $d$33-meter (APC International, Ltd.) that Allowi to measure piezoelectric coefficients in the range of 1–200 pace/V with an accuracy of 0.1 pal/V and ± 2% error. The temperature dependence of the piezoelectric coefficient was measured using a thermal chamber within the range of 293–357 K with temperature stabilization about 0.02 K. Laser-induced piezoelectric effect was studied using a cw Nd: YAG laser diode and its second harmonic generation. The power of this laser varied in the range of 200–400 mW. The photon energy at 532 nm lies above the edge of the absorption band of the studied crystals. The second and third harmonic generation studies were performed using Nd:YAG 10 ns pulse laser with 1064 nm wavelength, with maximal energy 80 mJ, pulse repetition frequency 10 Hz, beam diameter 8 mm, as the fundamental radiation source. The energy value was gradually tuned by Glan polarizer. The maximum energy density for the second harmonic was 2 100 J/m2. The energy density values of the incident fundamental radiation were measured with germanium detector, and its second harmonic with a photomultiplier with an interference filter with 5 nm half-width that transmits electromagnetic radiation at 532 nm wavelength.

## 3 Results and discussion

The crystal structure of the AgGaGe3Se8 crystal can be represented as the packing of the tetrahedra of selenium atoms around cationic atoms (Parasyuk et al. 2012; Reshak et al. 2013; Kityk et al. 2017). Silver atoms are located in the channels of the voids between these tetrahedra. In the first approximation of the closest atoms, the coordination surrounding of silver atoms can be considered as heavily distorted tetrahedra. The variants

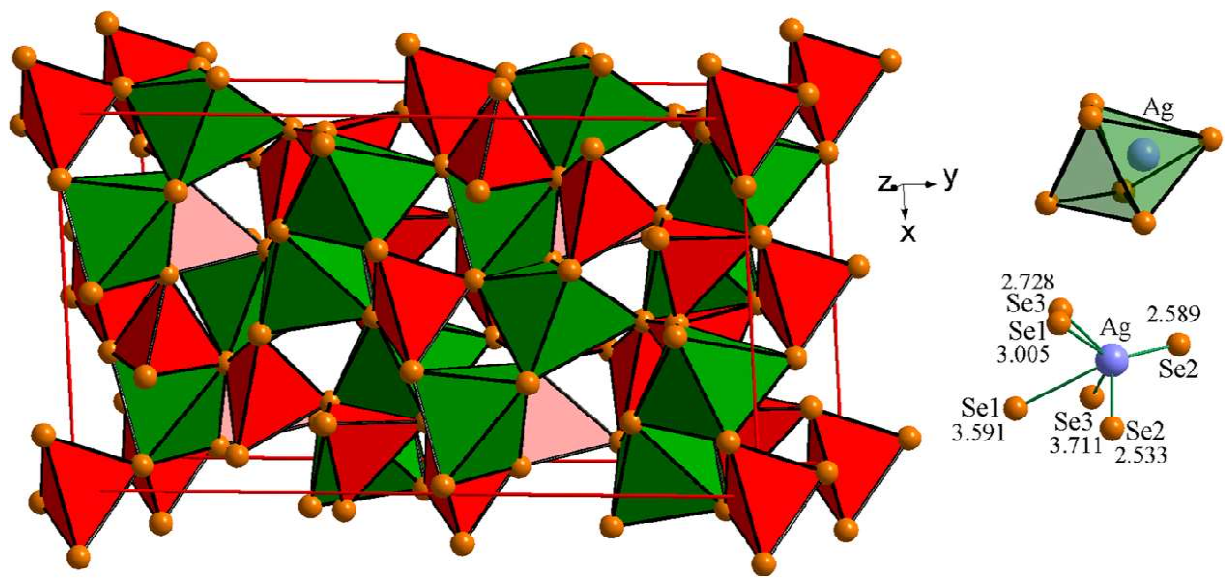

Fig. 1 Packing of the polyhedral of Se atoms surrounding Ga, Ge (red tetrahedral), Ag atoms (green octahedral) and inter-atomic distances Ag−Se in the structure of AgGaGe3Se8

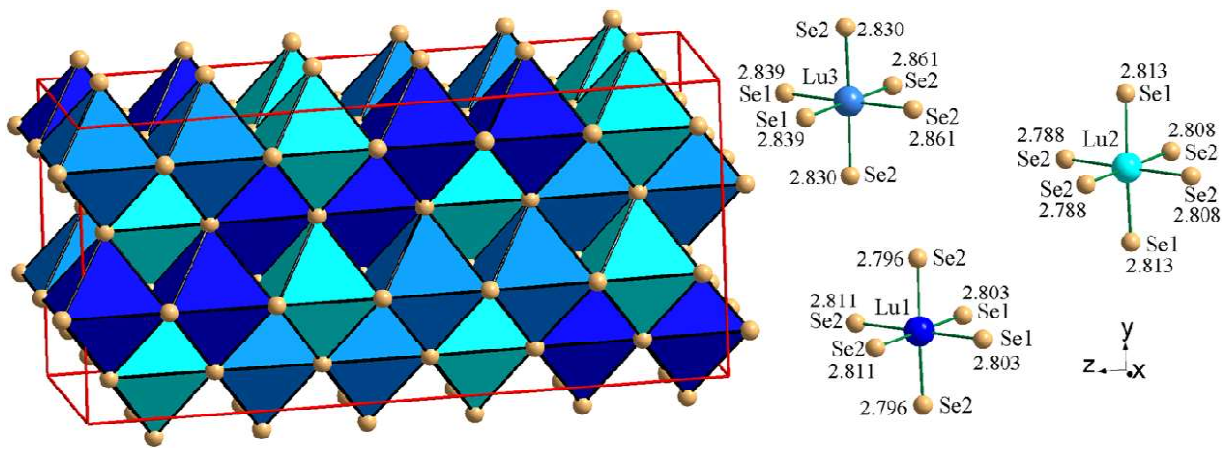

Fig. 2 Polyhedral representation and inter-atomic distances in the rt-Lu2Se3 structure (SG *Fddd*, No 70, *a* = 1.1251, *b* = 0.79806, *c* = 2.3877 (Ĺ)) (Folchnandt et al. 2004)

of the substitution with Cu (Kityk et al. 2013), In (Kuznik et al. 2016), Si (Kuznik et al. 2015), Sn (Kuznik et al. 2015) were considered in this assumption. The surrounding of silver atoms can also be represented as a distorted octahedron if we take into account the more distant atoms that are far beyond the sum of the atomic radii of silver (0.85 Ĺ) and selenium (1.91 Ĺ) (Fig. 1). The central atom is strongly displaced from the center of its octahedral environment. Taking into account that Lutetium atoms in binary selenides such as Lu2Se3 are in the octahedral surrounding (Folchnandt et al. 2004) (Fig. 2). Where the Lu–Se distances are in the range of 2.7–2.9 Ĺ, one can assume that the Lu atoms will occupy the crystallographic site *16b* statistically with Ag atoms forming solid solutions $Ag_{1-3x}Lu_xGaGe_3Se_8$ as a result of isoelectronic heteroatom substitution. The closeness of the radii and the small amount of lutetium atoms did not make it possible to clarify the position of the lutetium atoms but they are assumed to be more to the center of the octahedron than silver atoms. To explore a possibility to use $AgGaGe_3Se_8$: Lu in IR spectral range in the IR transmission spectra has been done. The corresponding results are shown in the Fig. 3. For convenience the spectra for undoped materials are given also in the Fig. 3 (El-Naggara et al. 2018).

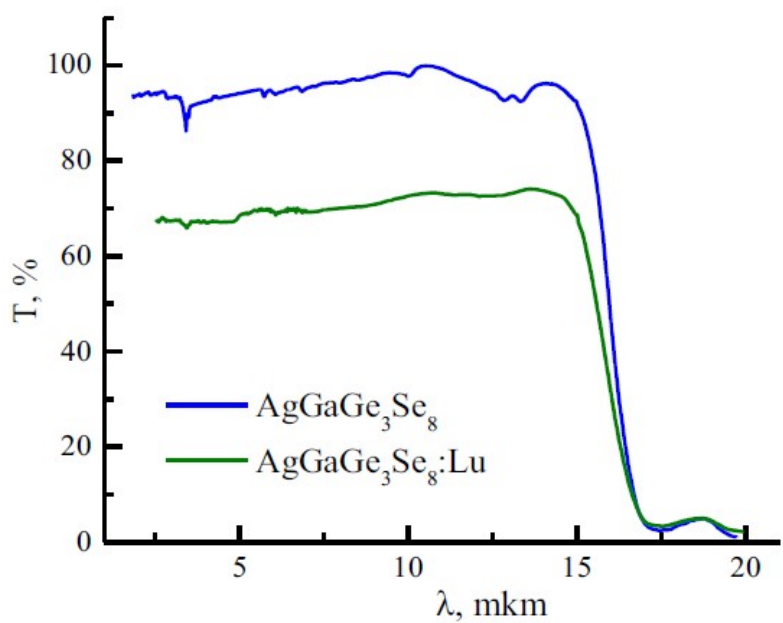

Fig. 3 Spectral distribution of the transmittance

IR transmission spectra were investigated to study the possibility of Rusing AgGaGe3Se8:Lu crystals to design the multifunctional optoelectronic devices for the IR region of the spectrum. The results are presented in Fig. 3. For comparison and analysis of the obtained results as well as the transmission spectra for the undoped AgGaGe3Se8 crystals taken from (El-Naggara et al. 2018) are also shown. The transmittance decreases in the whole transparency spectral range without significant changes of the spectra upon doping. The observed decrease in transparency is due to additional absorption and can be explained by assuming low-angular scattering of Light on inhomogeneities formed by the accumulation of charged impurities. A similar behavior was observed in Hg3In2Te6 crystals upon doping by gadolinium (Grushka et al. 2000). The potential of use semiconductors in optoelectronics depends on their band gap energy magnitude. In view of this, we explored the absorption spectra near the band energy gap (Fig. 4) and we have estimated the band gap energy (see Fig. 5).

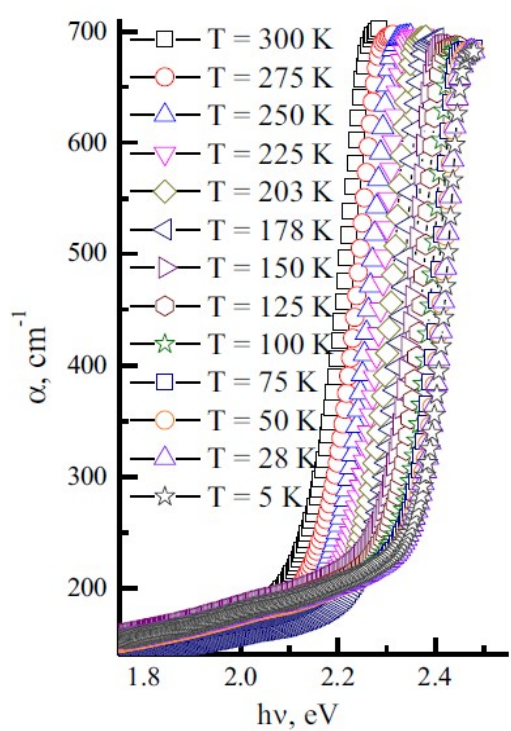

Fig. 4 Fundamental absorption edge of Baggage3Se8: Lu at different temperature

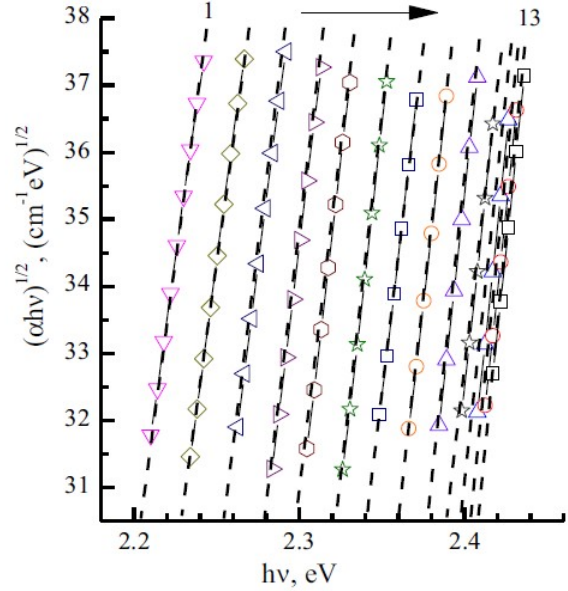

Fig. 5 The absorption spectra in the Tauc coordinates at different temperature: 1–300 K, 2–275 K, 3–250 K, 4–225 K, 5–203 K, 6–178 K, 7–150 K, 8–125 K, 9–100 K, 10–75 K, 11–50 K, 12–28 K, 13–5 K

The absorption coefficient was calculated by equation (Pankove 1975):

$$T = \frac{(1-R)^2 \exp(-\alpha d)}{1 - R^2 \exp(-2\alpha d)}, \qquad (1)$$

where $\alpha$ is the linear absorption coefficient; $d$ is the sample's thickness; $T = I/I0$—transmittance; $R$ is the reflection coefficient. The solution of Eq. (1) with respect to $\alpha$ allows obtaining an equation:

$$\alpha = \frac{1}{d} \ln \left\{ \frac{(1-R)^2}{2T} + \sqrt{\left[\frac{(1-R)^2}{2T}\right]^2 + R^2} \right\}. \qquad (2)$$

Spectral dependence of the calculated light absorption coefficient $\alpha$ at the fundamental absorption edge is presented in Fig. 4. The band gap energy was estimated by Tauc method (1974) according to the expression:

$$(\alpha h\nu)^{1/N} = f(h\nu), \qquad (3)$$

where N is the exponent that depends on the nature of the electronic transition responsible for the absorption (N = 1/2 for direct permitted transitions, N = 3/2 for direct forbidden transitions, N = 2 for indirect permitted transitions, N = 3 for indirect forbidden transitions). Extrapolation of the linear part of the graph obtained from Eq. (3) onto the energy axis defines the energy band gap of the sample. According to the experimental data (Lakshminarayana et al. 2012) and theoretical calculations (Reshak et al. 2013), AgGaGe3Se8 crystals have indirect band gap. The linear dependences in this work are obtained from the relation $(\alpha h v)^{1/2}$ versus $hv$, i.e. $N = 2$ which confirms indirect allowed transitions in the investigated crystals. The band gap value wass estimated by extrapolating straight lines to $(\alpha h v)^{1/2} = 0$ (Fig. 5).

Table 1 Band gap energy, Urbach's energy and steepness parameter at various temperatures

| Temperature (K) | $E_{gi}$, eV | $E_U$, meV | Steepness parameter |
|---|---|---|---|
| 290 | 2.06 | 111 | 0.233 |
| 275 | 2.08 | 107 | 0.221 |
| 250 | 2.10 | 99 | 0.217 |
| 225 | 2.12 | 96 | 0.202 |
| 203 | 2.15 | 91 | 0.192 |
| 178 | 2.18 | 86 | 0.178 |
| 150 | 2.20 | 83 | 0.156 |
| 125 | 2.23 | 77 | 0.140 |
| 100 | 2.25 | 75 | 0.115 |
| 75 | 2.26 | 71 | 0.091 |
| 50 | 2.28 | 69 | 0.063 |
| 28 | 2.286 | 68 | 0.035 |
| 5 | 2.290 | 68 | 0.006 |

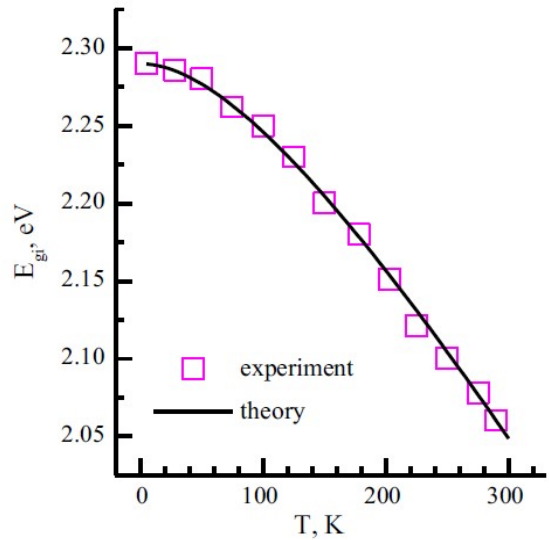

Fig. 6 Temperature dependence of the indirect band gap (*Egi*)

The temperature change of the indirect band gap energy is presented in Table 1 and the open points in Fig. 6. The dependence was fitted by empirical Varshni equation (Varshni 1967)

$$E_{gi}(T) = E_g(0) - \frac{\gamma T^2}{T + \beta}, \qquad (4)$$

where $E_g(0)$ is the band gap energy at 0 K, $\gamma$ is the rate of the temperature change of the band gap energy, $\beta$ is a temperature close to the Debye temperature. The solid line in Fig. 6 presents the best fit of the experimental and theoretical (Eq. 4) temperature dependence of the band gap variation. The calculated parameters

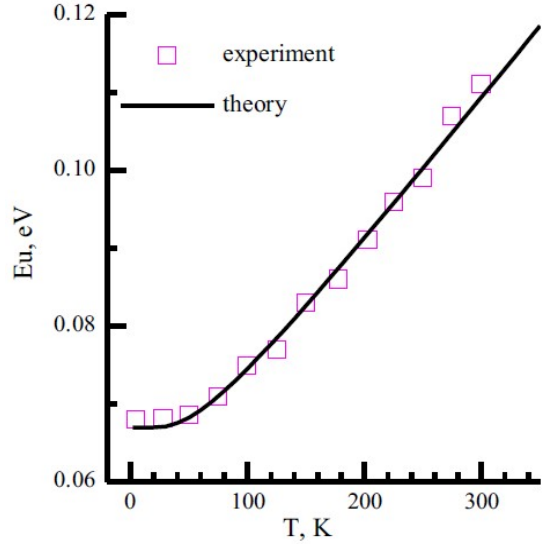

Fig. 7 Urbach energy as a function of temperature. The solid line is a fitting curve by Eq. (5)

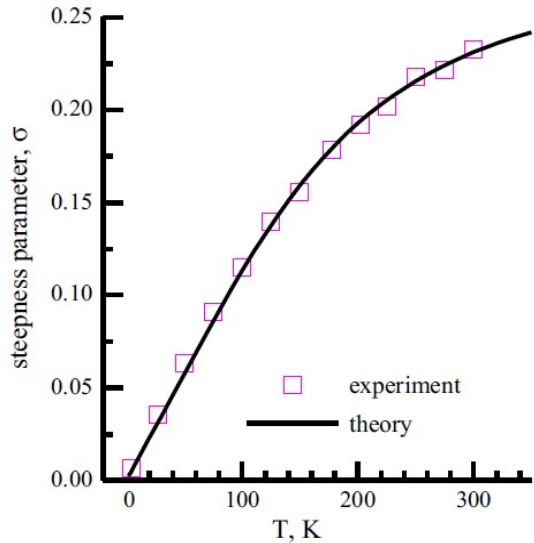

Fig. 8 Steepness parameter as a function of temperature. The solid line is a fitting curve by Eq. (6)

are $\gamma = 13.8 \cdot 10^{-4}$ eV/K, $\beta = 214$ K. The results agree well with the Debye temperature for the AgGaSe2 crystals which is 206.95 K (Hou et al. 2014). The absorption coefficient below the band edge is an exponential function of the Photon energy and obeys the following expression (Urbach 1953):

$$\alpha = \alpha_0 \exp\left(\frac{E - E_0}{E_U}\right), \qquad (5)$$

where $E$ is the photon energy, $E_0$ is the band gap energy near 0 K, $EU$ is Urbach's energy. $EU = kBT/\sigma(T)$ is defined as the inverse logarithmic slope of the absorption coefficient, and $\sigma(T)$ is the ssteepness parameter. The temperature changes of Urbach's energy and the slope parameters are presented in Table 1 and Figs. 7, 8. The increase of $EU$ with the temperature from 5 to 300 K is characteristic which is assumed to be caused by the increase due to the thermal ionization in the concentration of charged defects that were neutral at low temperature. At the same time, some contribution to this increase is made by additional ionization of the defect centers in the sample chich causes deviation from the periodicity of the potential by a random electric field created by the fluctuation of the concentration of charged defects (Mott and Davis 1971). In the case of significantly disordered surfaces of powders this contribution may be significant. The temperature dependence of Urbach's energy was used to evaluate the dominant mechanisms contributing to the blurring of the absorption edge. It was found that $EU$ can be modeled as an Einstein oscillator (Yang et al. 1995) which takes into account the contribution of dynamic (thermal) and static (structural and composition) disorder. According to this model, Urbach's energy can be expressed as:

$$E_U = [E_U(T) + E_U(X, C)] = A\left(\frac{1}{\exp(\Theta/T) - 1}\right) + B, \quad (6)$$

where $A$ and $B$ are constants associated with thermal, structural, and compositional disorder; $\Theta$ is the Einstein temperature which is related to the Debye temperature as $\Theta \approx l'\Theta_D$ (Abay et al. 2001). The first term of Eq. (6) corresponds to the contribution of electron–phonon interaction due to the Debye–Waller factor, and the second term is the standard deviation of atoms from the equilibrium sites, a structural disorder from a perfectly ordered lattice. The Best fit of the experimental results (open points) and Eq. (6) with the adjustable parameters $A$ and $B$ (solid line) is presented in Fig. 7. The adjustable parameter $A$ is equal to 30 meV and the parameter $B$—67 meV. Given the values of parameters A and B, structural and compositional disturbances ($B$) dominate because they contribute more to Urbach's energy than thermally induced disorder ($A$). The slope parameter $\sigma(T) = k\,T\,\Delta(\ln \alpha)/\Delta(h\nu)$ was calculated from the experimental data on the tails of the fundamental absorption edge (Fig. 8). The $\sigma(T)$ dependence is approximated in the entire studied temperature range by an expression for the absorption edge which is formed with the participation of electron–phonon interaction (Kurik 1971; Studenyak et al. 2014):

$$\sigma(T) = \sigma_0(2kT/h\nu_0)th(h\nu_0/2kT), \quad (7)$$

where $h\nu_0$ is the effective phonon energy which in most cases coincides with the phonons involved in the formation of the long-wave side of the absorption edge; $\sigma_0$ is temperaturein dependent but material-dependent parameter that is inversely proportional to the electron/exciton–phonon interaction constant $g$ by the formula $\sigma_0 = (2/3)g - 1$. Experimental data were fitted by Eq. (7) with $\sigma_0$ and $h\nu_0$ as adjustable parameters to estimate the phonon energy values associated with Urbach's tails. The best fit of the data is presented in Fig. 8 as solid curves. The approximation parameters were equal to $\sigma_0 = 0.28$, $h\nu_0 = 42$ meV. The effective phonon energy $h\nu_0$ is higher than the highest optical mode for AgGaSe2 and AgGaS2 crystals which are 276 cm$^{-1}$ (34 meV) (Miller et al. 1976) and 312 cm$^{-1}$ (39 meV), respectively

(Valakh et al. 2017). Higher $hv0$ value is related to the structure and composition disorders caused by cation substitution, cation vacancies, interstitial atoms, and deviation from stoichiometry. One of the main reasons for using chalcogenide crystals as laser materials is the large number of intrinsic non-stoichiometric cation defect states, energy positons of which are located within the band gap. They interact effectively with external laser radiation Fielding a great potential for improvement of optical polarizations (Kityk et al. 2013). The studdied crystal is noncentrosymmetric and, according to optical studies, highly defective, which makes it promising for use as a laser-operated piezoelectric material. In view of this, the possibility of using the material as a laser-operated piezoelectric module was tested. The origin of optically induced piezoelectricity is mainly optically induced trapping and rechargding of defect levels, as well as by some contribution of anharmonic photostimulated phonons (Narasimha et al. 2012). Piezoelectric coefficients were determined for different tensor components which Take into account the anisotropy of the corresponding components to find the maximum piezo electric constants. The piezoelectric coefficients of the AgGaGe3Se:Lu crystal at 302 K are $d11 = 3.4$ pm/V, $d22 = 13.6$ pm/V, $d33 = 29$ pm/V. Substantial anisotropy of three diagonal tensors reflects the high anisotropy of the studdied crystals. As the piezoelectric coefficient $d33$ has the highest value, its behavior for AgGaGe3Se:Lu was further investigated. The variation in the fundamental piezoelectric tensor coefficient with increasing temperature from 293 to 343 K and subsequent cooling is presented in Fig. 9. This piezomodule is changed almost linearly versus temperature. As the temperature increased by 50 K, the value of $d33$ is increased by 82%. The possible reason for this growth is the decrease in the bandgap energy and the corresponding increase in free charge carriers effectively interacting with the phonon and electron subsystems (Nouneh et al. 2006). Another reason for the change may be the thermal expansion of the lattice parameters which in turn leads to spatial redistribution of charge density and increase in noncentrosymmetry. The changes are completely reversible upon cooling. The obtained results agree well with (Kityk et al. 2017) according to which significant changes in piezo electric coefficients are related to the distortions of the basic structural units that determine the acentricity of the charge density. The piezoelectric properties are sensitive not only to temperature but also to laser irradiation. According to Kityk et al. (2013) and Williams et al. (2006), there are two main effects that determine the transport of charge carriers under laser irradiation, the change of piezoelectric resistance due to slight changes in the band gap caused by laser stimulation, and the change of charge density distribution which forms the so-called polar piezopotential. The value of this potential depends on the polarization ability and the concentration of intrinsic defect states. The irradiation changes of the static dipole moments which are fundamental to the effects are described by the third rank tensors. At the same time, contrary to other compounds, chalcogenide materials exhibit significant anharmonic Photon contributions (Shpotyuk et al. 1997; Kityk 2004) that affect the piezoelectric properties of the crystals. The results of the investigation of optically induced piezoelectric effect for the AgGaGe3Se8: Lu crystals are presented in Fig. 10. Turning on the continuous-wave 400 mW laser at 532 nm which is close to the band gap energy of the crystals, leads to significant decrease in the piezoelectric coefficient from 28.7 to 15.6 pm/V is observed. This can be explained by the redistribution of the local charge density by an external optical laser field. When the laser is turned off, the piezo electric module values return to their initial values. The relaxation of photoinduced chan ges lasts for tens of seconds (Fig. 10) confirming the involvement of localized intra-band metastable traps in the observed effects. The studied samples can withstand a large number of such cycles of irradiation, and no irreversible changes in the properties of the crystals were observed after 10 cycles. We believe that this effect can be used to form laser-controlled piezoelectric devices. The conversion of IR frequency of laser sources (frequency modulation) is one of the main tasks of optoelectronics. Therefore we studied the second harmonic generation (the second-order optical effect) which is governed by the third-rank tensors. SHG measurements were performed by the traditional Kurtz−Perry powder technique (Kurtz and Perry 1968). It should be noted that the Kurtz Perry method allows only a relative evaluation of the efficiency of non-linear optical transformations using only the powder- like of a non-linear optical material. This method makes possible the testing of New promising NLO materials using the approximation that the powder particles are considered like single crystals with average dimensions of ~ 10···120 $\mu$ m that are randomly orientem in space. The grain sizes of the powder obtained by mechanical crushing were estimated using SEM (FEI VERSA 3D) (Fig. 11). It should be emphasized that due to the powderlike grain structure, the contribution to the optical and non-linear optical constants will be principally different with the respect to the bulk flawless materials surfaces. The powders for SEM measurements were mounted (glued using the Ag containing paint) on the Al substrates. The studied specimens put onto the surface of the paint and they have been dry to remove the liquid components of the paint. We have used relatively low temperatures of drying within the 25−50 °C. Prepared materials put on the Al tables, which simultaneously allow to varying the orientation of the sample. The position of the measuring table was tuned. The principal goal of the studies was an observation of the structure of the titled materials and evaluations of the grains sizes and shapes. Figure 11 presents selected photographs of tested powders observed at magnifications from 100 up to 650 xs. Observations shows that the grain size of AgGaGe3Se8 powder is smaller than Lu doped AgGaGe3Se8. The size uniformity is also greater. Some grains of powder were set at an angle or covered by others, making it difficult to measure the size and shape. The shape of the particles generally can be described as irregular. Their linear dimensions are not the same in different directions. For the AgGaGe3Se8 material, grains of about 10 to about 120 $\mu$m have been detected. In the case of Lu-doped material, these sizes ranged from about 18 to as much as 270 $\mu$m. It should be mentioned that in the Lu-doped material, the amount of grains larger than 100 $\mu$m was much larger. In addition, at higher magnifications up to 350,000 times, the presence of nano-size dust in the research samples was detected. In the case of the AgGaGe3Se8 material in the volume dust there were more. This dust was created as a result of mechanical grinding. The grain size has a very significant effect on the intensity of the second harmonic signal. The proportion of smaller powder grains gives higher second harmonic signals. It is associated with a larger surface interacting with laser radiation. The results of the nonlinear optical SHG, photoconductivity and laser stimulated piezoelectricity are presented for the AgGaGe3Se8 and AgGaGe3Se8: Lu crystallutes. The intensity of the second harmonic generation in the doped crystals is lower than that in the undoped samples. In our opinion, this can be caused both by the change of the band gap energy and the changes in effective phonon contributions. The doping by Lu Leeds to additional absorption at 532 nm and to additional contribution to the hyperpolarizabilitines with different signs. For the powders bulk-like and reconstructed caused by a effect mesoscopic contribution may give a comparable contribution to the SHG in the enhanced optical susceptibilities selected regime (Figs. 12, 13, 14).

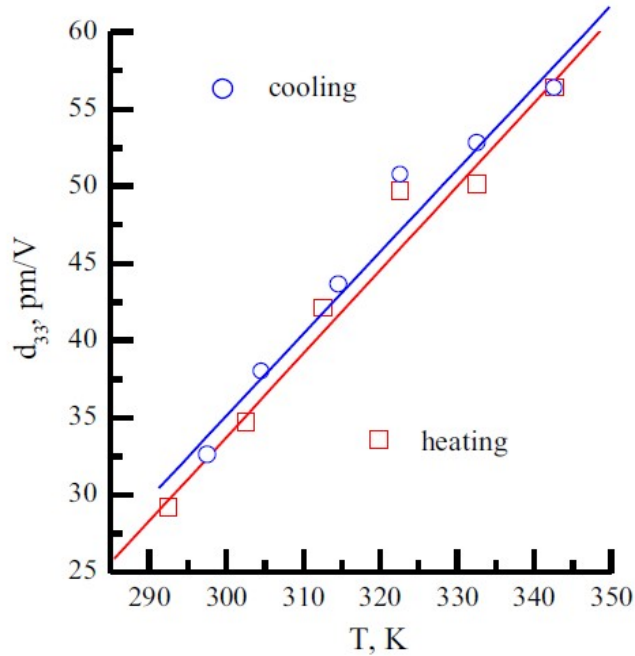

Fig. 9 Temperature dependence of the principal piezo electric coefficient for the AgGaGe3Se8: Lu samples

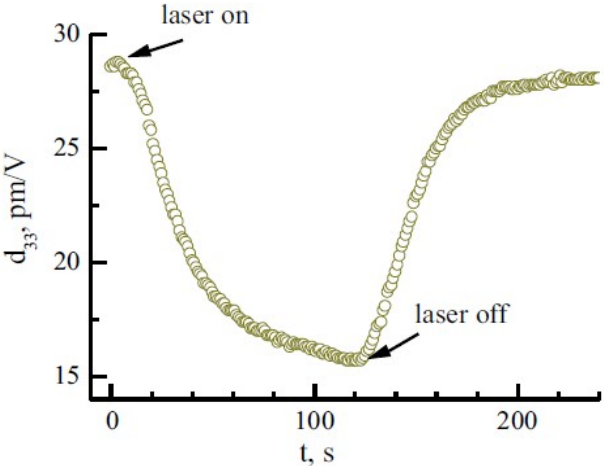

Fig. 10 Time kinetics of the piezoelectric photo-induced changes

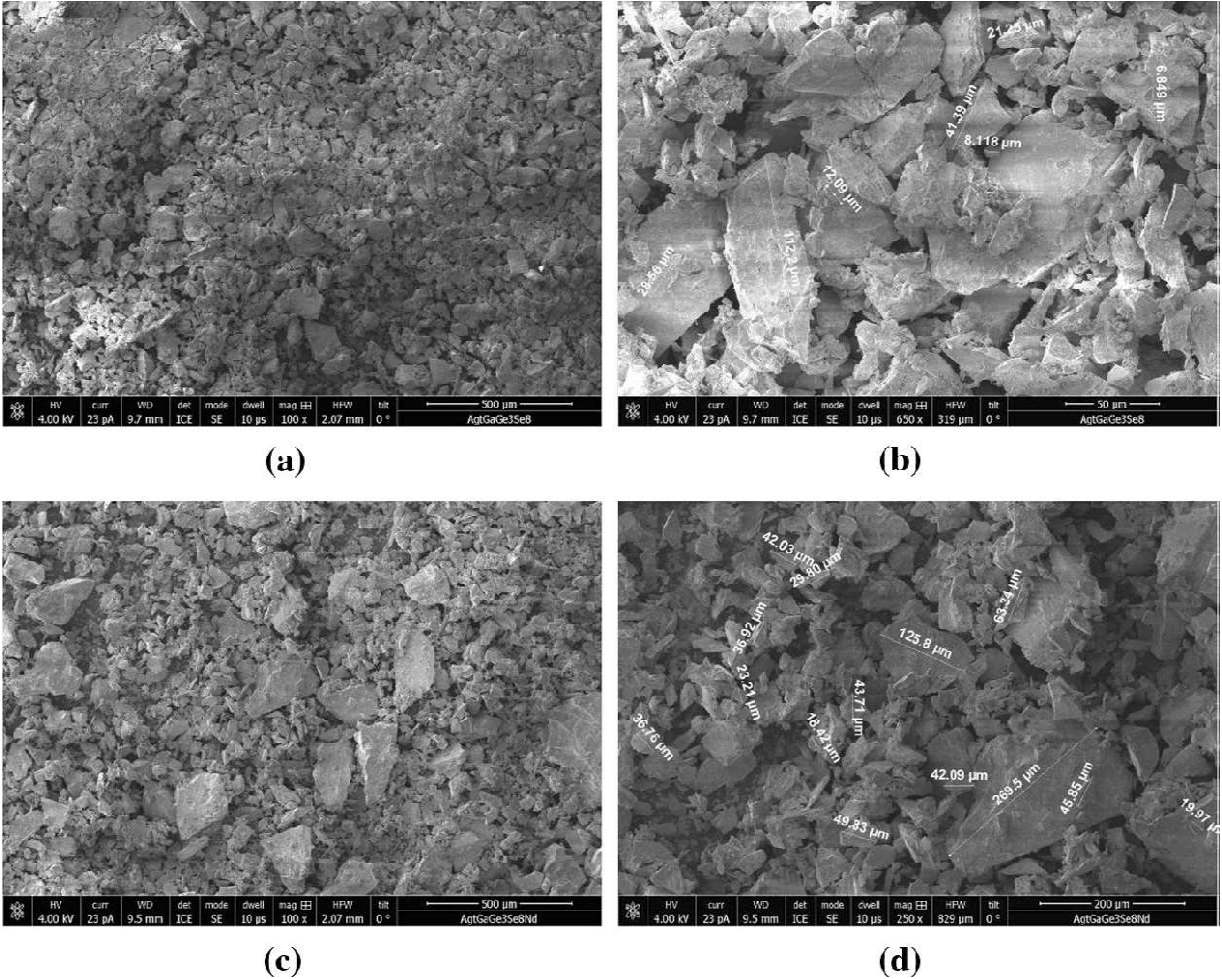

(a)                  (b)

(c)                  (d)

Fig. 11 Evaluation of particle size by SEM: **a**, **b** AgGaGe3Se8; **c**, **d** AgGaGe3Se8:Lu

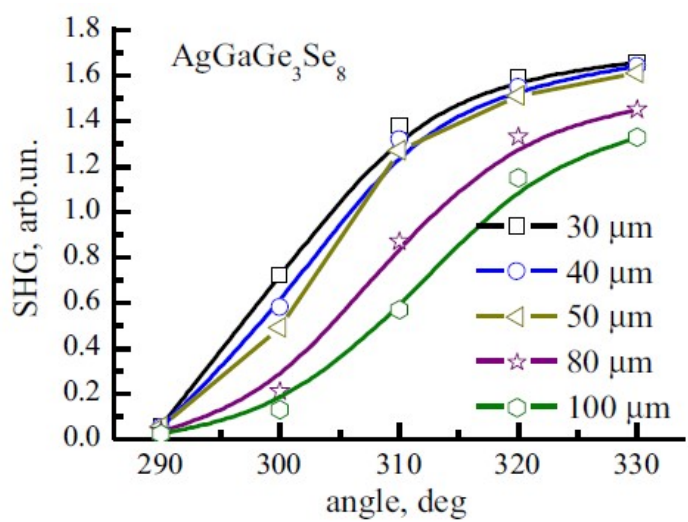

Fig. 12 Angle-dependent second harmonic generation for AgGaGe3Se8

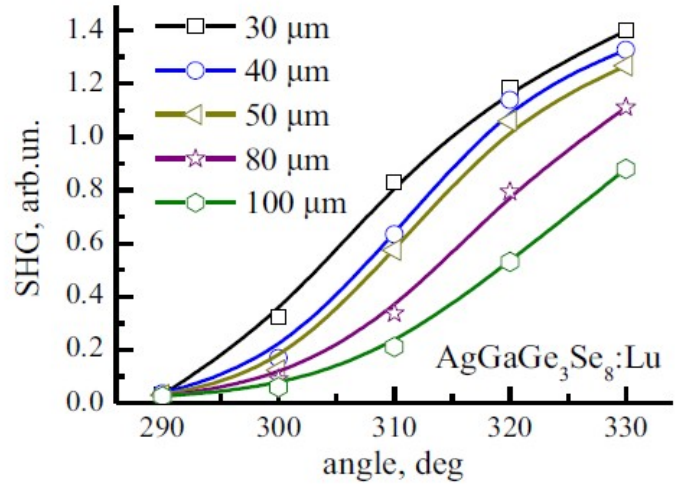

Fig. 13 Angle-dependent second harmonic generation for AgGaGe3Se8:Lu

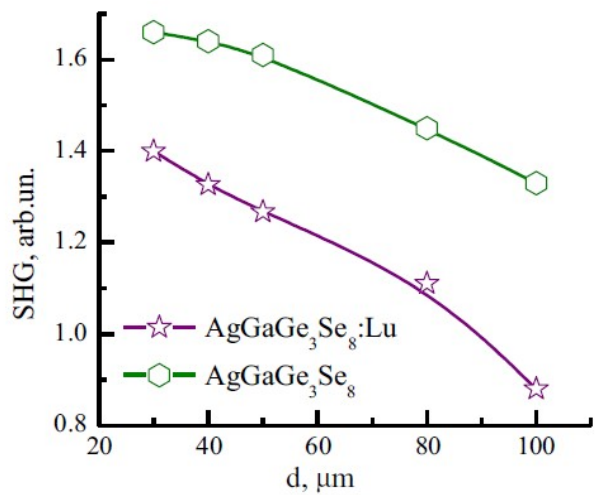

Fig. 14 SHG efficiencies depending on the grain size

## 4 Conclusions

A complex study of Lu-doped AgGaGe3Se8 crystals and comparison with undoped samples was performed by spectroscopic, piezoelectric, nonlinear optical, photonductivity methods in reflecged geoemtry. Particular role play surface morphological sheets. To study the possibility of using AgGaGe3Se8: Lu crystals in the infrared spectrum to design multifunctional materials for optoelectronic devices, IR transmittance spectra and the spectral distribution of the absorption coefficient in the region of the fundamental absorption edge were investigated. The AgGaGe3Se8:Lu crystals were found to be indirect-gap materials. The band gap energy in the temperature range 5−300 K was determined and has show principal role of Photon subsytem. It is crucial that the intensity of the second harmonic generation in the doped crystals is lower than that in the undoped samples. In our opinion, this can be caused both by the change of the band gap energy aand the changes in effective phonon contributions. The grain sizes and related susceptibilities dopin by rare earth significantly change. The doping by Lu leads to additional absorption at 532 nm and to additional contribution to the hyperpolarizabilitines with different signs. For the powders bulk-like samples principal role begins to play effective mesoscopic contribution may give a comparable contribution to the

SHG as the bulk crystals. The parameters of Urbach's rule were investigated to evaluate the dominant mechanisms that contribute to the tailing of the fundamental absorption edge. The studies of temperature and laser-induced piezoelectric effects and SHG indicate that AgGaGe3Se8: Lu is a promising multifunctional material for mid-IR applications.

**Acknowledgements**

The presented results are part of a project that has received funding from the European Union's Horizon 2020 research and innovation program under the Marie Skłodowska-Curie Grant Agreement No. 778156. K.I.V, and J.J., acknowledge support from resources for science in the years 2018−2022 Granted for the realization of international co-financed Project Nr W13/H2020/2018 (Dec. MNiSW 3871/H2020/2018/2).